# Deconfined quantum critical point lost in pressurized SrCu$_2$(BO$_3$)$_2$

Jing Guo,[1,5,*] Pengyu Wang,[1,2,*] Cheng Huang,[3,*] Bin-Bin Chen,[3] Wenshan Hong,[1,2] Shu Cai,[4] Jinyu Zhao,[1,2] Jinyu Han,[1,2] Xintian Chen,[1,2] Yazhou Zhou,[1] Shiliang Li,[1,2,5] Qi Wu,[1] Zi Yang Meng,[3,†] and Liling Sun[1,2,4,5,‡]

[1]*Beijing National Laboratory for Condensed Matter Physics and Institute of Physics,*
*Chinese Academy of Sciences, Beijing 100190, China*
[2]*University of Chinese Academy of Sciences, Beijing 100190, China*
[3]*Department of Physics and HKU-UCAS Joint Institute of Theoretical and Computational Physics,*
*The University of Hong Kong, Pokfulam Road, Hong Kong SAR, China*
[4]*Center for High Pressure Science & Technology Advanced Research, 100094 Beijing, China*
[5]*Songshan Lake Materials Laboratory, Dongguan, Guangdong 523808, China*
(Dated: October 30, 2023)

In the field of correlated electron materials, the relation between the resonating spin singlet and antiferromagnetic states has long been an attractive topic for understanding of the interesting macroscopic quantum phenomena, such as the ones emerging from magnetic frustrated materials, antiferromagnets and high-$T_c$ superconductors [1]. SrCu$_2$(BO$_3$)$_2$ is a well-known quantum magnet, and it is theoretically expected to be the candidate of correlated electron material for clarifying the existence of a pressure-induced deconfined quantum critical point (DQCP) [2, 3], featured by a continuous quantum phase transition, between the plaquette-singlet (PS) valence bond solid phase and the antiferromagnetic (AF) phase. However, the real nature of the transition is yet to be identified experimentally due to the technical challenge. Here we show the experimental results for the first time, through the state-of-the-art high-pressure heat capacity measurement, that the PS-AF phase transition of the pressurized SrCu$_2$(BO$_3$)$_2$ at zero field is clearly a first-order one. Our result clarifies the more-than-two-decade-long debates about this key issue, and resonates nicely with the recent quantum entanglement understanding that the theoretically predicted DQCPs in representative lattice models are actually a first-order transition [4–8]. Intriguingly, we also find that the transition temperatures of the PS and AF phase meet at the same pressure-temperature point, which signifies a bi-critical point as those observed in Fe-based superconductor and heavy-fermion compound, and constitutes the first experimental discovery of the pressure-induced bi-critical point in frustrated magnets. Our results provide fresh information for understanding the evolution among different spin states of correlated electron materials under pressure.

## INTRODUCTION

The deconfined quantum critical point (DQCP) is a concept to describe the continuous phase transition between two spontaneous symmetry-breaking phases in the correlated electron material at zero temperature [9, 10]. It is characterized by the absence of confinement – the state of elementary excitations carries fractionalized quantum number and interacts via emergent gauge field – different from those as predicted by the conventional phase transitions [11–13]. The concept of the DQCP receives widespread attention because it gives rise to exotic states of matter, along with the Berezinskii-Kosterlitz-Thouless (BKT) transition [14, 15], anyon condensations [16, 17] and those in topological insulators and high-temperature superconductors, to challenge the conventional understanding of matter within the paradigm of Landau-Ginzberg-Wilson (LGW) where symmetries and their spontaneous breaking are the dominated factors [11, 12]. The concepts of fractionalization and emergent gauge fields in DQCP also have potential applications in quantum computing and information perspective.

In the past two decades, investigations on DQCP have been an active subject. Enormous efforts have been made to explore various theoretical models and experimental systems to realize the DQCP and its associated consequences [18–30]. However, a key question about whether the proposed DQCP models really host a continuous phase transition corresponding to a true conformal field theory (CFT) remains controversial and is still under intensive debates. Previous results suggest that the DQCP is realized in some 2D quantum spin or interacting Dirac fermion lattice models [18, 30] and the transition of two different spontaneous symmetry breaking phases indeed appears to be continuous, but the finite-size scaling is not consistent with regular scaling ansatz [21, 23, 31], and the extracted scaling dimensions from the numerical simulations are incompatible with later evaluations based on conformal bootstrap [32]. The controversy is more clearly revealed very recently that the DQCP "fails" a series of general standards from quantum entanglement perspective that all CFTs are expected to meet [4–7, 33, 34], making the nature of the DQCP more enigmatic at the present after two decades of debating.

At the experimental frontier, material realizations and detections of the DQCP are equally active. In quantum magnets, it was found that the layered material, SrCu$_2$(BO$_3$)$_2$, could be a better candidate for realizing DQCP due to its unique crystal structure [2, 3, 35–42]. As shown in Fig. 1 (a) and (b), Cu$^{2+}$ ions (in 3d$^9$ configuration) in monolayer SrCu$_2$(BO$_3$)$_2$ form a dimerized singlets (DS) state at ambient pressure, which couples the Cu$^{2+}$ with a strong intradimer antiferromagnetic coupling $J'$ and a weaker inter-dimer coupling $J$. The 2D network of the Cu$^{2+}$ ions resembles the famous Shastry-Sutherland (SS) lattice [43], as shown in Fig. 1 (b). Application of pressure ($P$) can change the lattice constants of the material and consequently alter the ratio of $J$ over $J'$. The previous works established the empirical relations of $J'(P) = (75 − 8.3P)$ K and $J(P) = (46.7 - 3.7P)$ K, with $P$ being in GPa, based on the data of high-pressure heat capacity measurements [2, 3]. By applying hydrostatic pressure above 1.8 GPa, $SrCu_2(BO_3)_2$ enters a plaquette-singlet VBS phase (Fig. 1 (c)). Upon further increase the pressure to above 3 GPa, the PS phase transforms to an AF phase [3] (Fig. 1 (d)). However, due to the technique limitations in the practical high-pressure measurements, the truly important information about the nature of the PS to AF transition remains unknown.

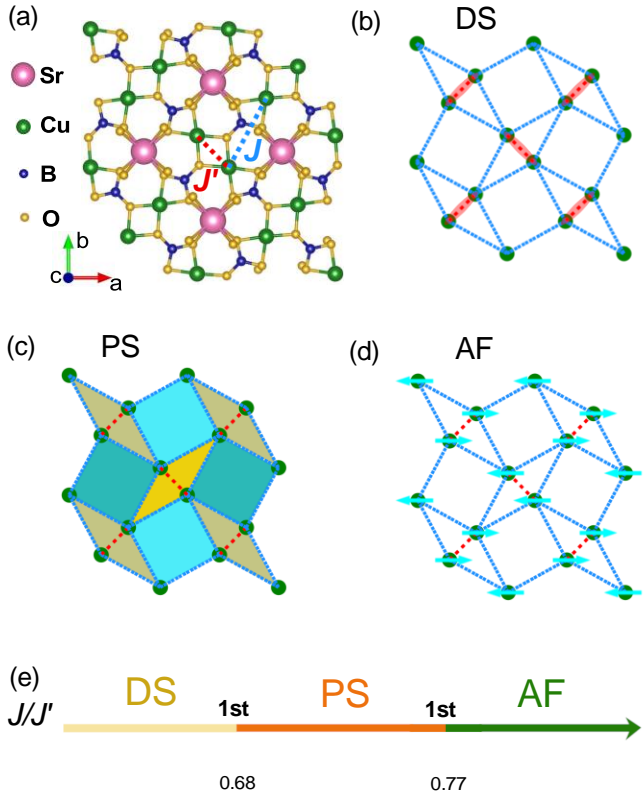

FIG. 1. **$SrCu_2(BO_3)_2$ structure, Shastry-Sutherland lattice and the related ground-state phase diagrams.** (a) Single layer of $SrCu_2(BO_3)_2$, where the intradimer $Cu^{2+}$-$Cu^{2+}$ interaction is labeled by $J'$ while the interdimer interaction is labeled by $J$. (b) Dimerized singlet (DS) state formed by $Cu^{2+}$ in $SrCu_2(BO_3)_2$, in which $J'$ is stronger than $J$ (in the Shastry-Sutherland (SS) lattice), at the ambient pressure. (c) (d) The emerging states, PS and AF as $J/J'$ increases. Note that in the PS phase only half of the plaquettes, either the yellow&beige or teal&cyan filled polygons, form singlets and breaks the symmetry of SS lattice. In the AF phase, the on-site spin rotational symmetry is broken. (e) Phase diagram of SS model by ED and DMRG calculations [38, 44], where the transitions are all first-order.

Recent experimental works show that application of magnetic field perpendicular to the ab-plane can trigger the transitions of the $SrCu_2(BO_3)_2$ material [35, 36], while the material exhibits a proximate critical yet first-order transition from the PS phase towards the AF phase [36] as the magnetic field is enhanced. Moreover, it was further hypothesized that this weakly first-order transition could eventually connect to a true DQCP or even a quantum spin liquid (QSL) phase close to zero field along the pressure axis [36].

To clarify this key puzzling issue of whether the PS-AF transition is a beyond-LGW DQCP or a LGW-allowed first-order transition experimentally (see Fig. 1 (e)) and to shine light on the overall ground-state phase diagram of the spin-1/2 SS model [38, 44–49], and further to clarify whether the material $SrCu_2(BO_3)_2$ could host an intermediate quantum spin liquid phase between the AF and PS phases [50–52], in this work, we perform high-pressure heat capacity measurements by using the state-of-art technique, which allows us to measure the sample in a hydrostatic pressure environment for pressures continuously tuned to above 3 GPa and low temperature down to 0.4 K (the details can be found in Supplementary Information-SI).

### TEMPERATURE DEPENDENCE OF HEAT CAPACITY AT DIFFERENT PRESSURES

As shown in Fig. 2, the plots of the temperature dependence of heat capacity versus temperature ($C/T$) displays a hump below 1.8 GPa as the temperature is reduced (Fig. 2 (a) and (b)), the maximum of which is associated with the formation temperature ($T_{DS}$) of the dimer singlet phase [37–39, 41, 42]. Upon increasing pressure to 2.1 GPa, there are two peaks appearing at low temperature. The high temperature one is related to the onset transition temperature from the paramagnetic (PM) phase to the PS liquid (PSL) phase ($T_{PSL}$), and the low temperature peak is associated with the transition temperature of PSL-PS phase ($T_{PS}$). These two phases present in the range of 2.1-2.6 GPa (Fig. 2 (c)-(e)). At pressure about 2.7 GPa, an AF phase appears at temperature slightly lower than $T_{PS}$ (Fig. 2 (f)). It seems like that the three phases, AF, PS and PSL phases are compatible (Fig. 2 (f)). Such a feature can be observed until 2.75 GPa (Fig. 2 (g)). At 2.9 GPa and above, the PS phase no longer exists (the reason will discuss below) but the AF phase prevails (Fig. 2 (h)). A hump feature is also observed at the temperature higher than $T_{PSL}$, we define the temperature at the maximum hump as the onset temperature of the AF liquid state ($T_{AFL}$), below which the spins start to establish the effective AF interactions but are not ordered yet [3]. The true AF long-range order is established below the $T_{AF}$ peak, as shown in Fig. 2 (h) for 2.9 GPa. We repeat the experiments with a new sample cut from the different batches and observe the reproducible results (see SI).

### ESTABLISHMENT OF COMPLETE PRESSURE-TEMPERATURE PHASE DIAGRAM

We summarize our experimental results in the phase diagram (Fig. 3). All the data shown in the main panel are from the heat-capacity measurements, in which the solid and half-filled markers are the data obtained from two samples separately in this study, while the open markers are the results from our previous study [3]. All data measured on the different samples are well consistent with each other. There are three regimes in the phase diagram: below 1.8 GPa, the ground state of the SrCu$_2$(BO$_3$)$_2$ sample holds DS phase (the left regime); at pressure about 1.8 GPa, the system enters the PS phase (the middle regime); further compression to about 2.7 GPa, the AF phase sets in and stabilizes up to 4 GPa. To understand the details on the PS-AF phase transition, we zoom in the phase diagram near the boundary of these two phases (see the inset of the main panel). A coexistence of PS and AF phases is observed in a very narrow pressure range. To find the precise pressure point for the PS-AF phase transition, we extrapolate the plot of temperature versus pressure for the two phases near the boundary (see red dash and green dash lines), which gives rise to an intersection located at 2.78 GPa. This intersection marked by a red star is defined as the critical pressure for the PS-AF phase transition. It is apparent that pressure above 2.78 GPa, the PS phase completely transforms to the AF phase. Our results reveal that the transition from the PS to the AF phases is the first order transition, which gives the decisive answer that the lattice translation symmetry breaking PS phase evolves into the spin rational symmetry breaking AF phase. Such a scenario can be supported by the plots of heat capacity versus temperature measured at 2.7 GPa and 2.75 GPa (Fig. 2 (f) and (g)) , where the $C/T$ (T) at $T_{AF}$ is getting higher than that at $T_{PS}$ as the pressure approaches the AF regime. These changes are in concordance with the well-known behavior of a first-order transition.

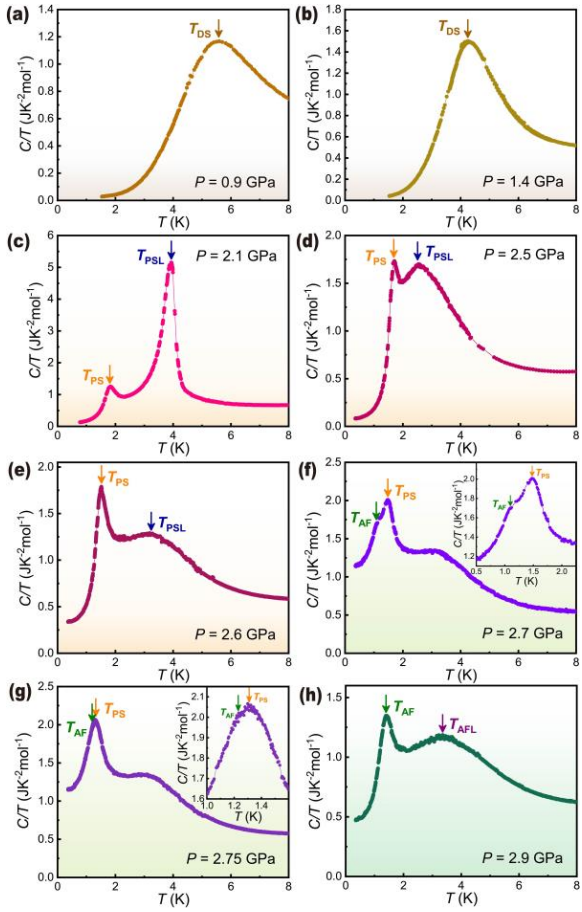

FIG. 2. **The results of temperature dependence of high-pressure heat capacity.** (a) and (b) are the plots of $C/T$ versus $T$ measured at 0.9 and 1.4 GPa, which show the typical characteristic of the dimmer singlet (DS) phase. The maximum of the hump $T_{DS}$ signifies the formation temperature of the DS phase. (c)-(e) display the pressure-induced two transitions upon cooling measured at 2.1, 2.5 and 2.6 GPa. The high temperature peak is related to the plaquette singlet liquid (PSL), while the low temperature peak is associated with the plaquette singlet phase (PS). (f-g) are the results obtained at 2.7 and 2.75 GPa, illustrating the emergence of the antiferromagnetic (AF) phase, close to the first order transition between the PS and AF phases. The insets show the transitions at low temperatures for better view. (h) presents the results of $C/T$ versus $T$ measured at 2.9 GPa, which exhibits an AF ground-state below $T_{AF}$ and crossover temperature $T_{AFL}$ from an antiferromagnetic liquid phase to paramagnetic phase at high temperature.

It is worth noting that the observation on the narrow overlap of the PS and AF phases should come from the inhomogeneity of the pressure, because SrCu$_2$(BO$_3$)$_2$ is a pressure-sensitive material [53–56]. A slightly difference of the pressure environment (the pressure at the center and the edges of the sample) may influence the detected critical points of the two phase transitions. Fortunately, this 'by-product' pro-vides us a way to determine the critical pressure/temperature point for the transition of PS-AF phases, which meets at a nearly same pressure/temperature point and is defined as the bi-critical point in the phase transition theory. The observation of the bi-critical transition between PSL-PS and AFL-AF phases in SrCu$_2$(BO$_3$)$_2$ is reminiscent of what have observed in the Fe-based superconductor Ca$_{0.73}$La$_{0.27}$FeAs$_2$ [8], in which the sample also undergoes a first order transition from the AF phase to a superconducting (SC) phase, accompanied by PM-AF and PM-SC transition at a bi-critical point. A similar observation has also been made in heavy fermion compound YbAgGe [57].

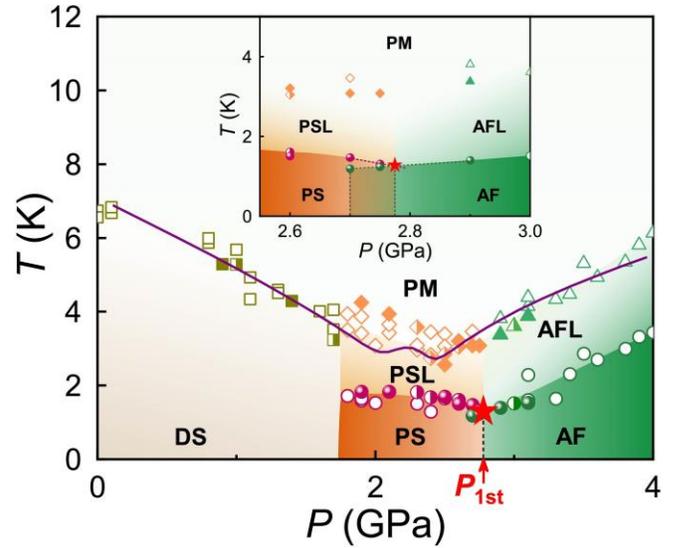

FIG. 3. **The complete $P$-$T$ Phase diagram of SrCu$_2$(BO$_3$)$_2$.** The acronyms PM, DS, PSL, PS, AFL, and AF stand for paramagnetic, dimer singlet, plaquette singlet liquid, plaquette singlet, antiferromagnetic liquid, and antiferromagnetic phases, respectively. $P_{1st}$ represents the pressure of the first order transition. The solid and half-filled markers are the data obtained from two samples separately in this study, while the hollow ones are the data obtained from our previous study [3], all of which is in good agreement. The inset zooms in the transition near the boundary of PS and AF phases. The red star is the critical pressure point of 2.78 GPa, which is determined by an extrapolation of onset temperatures of the PS and AS phases (see red and green dash line). The solid line which goes through the data points of $T_{DS}$, $T_{PSL}$ and $T_{AFL}$ (the crossover scale of the AF correlation, denoted by the green triangles), are from an ED calculation on the SS model with functions of $J'(P)$ and $J(P)$ mentioned above [3].

**PRESSURE DEPENDENT GAP OF DS, PS AND AF PHASES**

Since the gap value extracted from low temperature fit to $C/T$ can further diagnose the nature of the phases and the phase transition, we fit the low-temperature data based on the following form [3]: $C/T = a_0 + a_1 T^2 + (a_2/T^3)\exp(-\Delta/T)$, where $\Delta$ is the activation gap and $a_0$, $a_1$, and $a_2$ fitting parameters. Figure 4 shows the extracted gap by fitting $C(T)/T$ to an exponential form plus terms accounting for the heater, wires and phonons, together with the data reported previously [2, 3, 35]. Our results reveal that both DS and PS phases are gaped, which are in good agreement with both of the experimental and theoretical results [2, 3, 37–39, 41, 42]. We also observed a large reduction in gap at the boundary of the PS-AF phases. The clear drops in gap at the boundaries of the DS-PS and PS-AF phase transitions further confirm that these transitions are the first order ones. By the same method, we also fit the low-temperature data measured at pressure larger than 2.9 GPa, and find that the plots of $C/T$ versus $T$ cannot resolve an activation gap but render a power-law decay with temperature. This is the hallmark of the existence of the AF phase based on the gapless Goldstone modes [58]. As a result, we propose that SrCu$_2$(BO$_3$)$_2$ transforms to an AF phase completely via a first-order transition at 2.78 GPa.

**DISCUSSION**

The observation of the pressure-induced first-order transition between the plaquette singlet phase and the antiferromagnetic phase in the SrCu$_2$(BO$_3$)$_2$ delivers a clear message that the DQCP is lost in this pressurized material. In the recent theoretical developments of several important DQCP lattice models, both quantum spin models [5–7, 34] and inter- acting Dirac fermion models with Kane-Mele and plaquette interactions [4, 33], one consistently finds that the seemingly continuous transitions therein, irrespective from VBS to AF phases, or from quantum spin Hall (QSH) to SC phases, where both sides are spontaneously symmetry-breaking phases and the transitions are tuned by a single parameter, cannot be compatible with more fundamentally "first principle" tests as a continuous transition with CFT description. These tests in-clude the conformal bootstrap bounds of critical exponents for emergent continuous symmetry and the positivity requirement of the entanglement entropy [4–7, 32–34, 59]. The messages from these analyses imply that, the DQCPs with their present lattice model realizations are either first-order transitions or some other more complicate scenarios, such as multicritical point and complex fixed points, which at the moment one does not have controlled theoretical framework to calculate their precise properties [24, 31, 60].

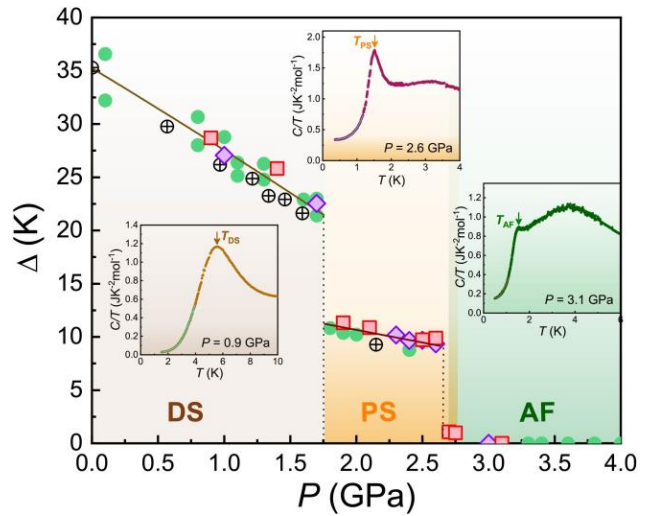

FIG. 4. **Pressure dependence of gap extracted from low-temperature fits to $C/T$.** The red squares and purple diamonds are the data obtained from this work and the rest symbols are the data from Refs. [2, 3]. The three insets are the plots of $C/T$ versus temperature measured at the three typical pressures with the fitting details in the low temperature.

Our experimental results in this study, therefore, come as a great relief for that the PS-AF transition in pressurized SrCu$_2$(BO$_3$)$_2$ at zero field and the ground-state is a first-order one, which puts the final piece of the puzzle onto the phase diagram and clarifies the two-decades-long debates on DQCPs of SrCu$_2$(BO$_3$)$_2$ and related models. Our results also resonate nicely with the recent quantum entanglement understanding that the proposed DQCPs in 2D quantum spin or interact- ing Dirac fermion models cannot be described by CFT [4– 7, 33, 34].

Another interesting observation in this study is that the $C/T$ data measured at very low temperature (roughly the $a_0$ in the gap fitting function) exhibits an enhanced value at the temperature lower than the bi-critical point (see Fig. S3 in the SI). This might suggest there exist enhanced fluctuations in the vicinity of the bi-critical point and imply that, if one could further suppress the bi-critical point by competing interactions or geometry frustrations, a DQCP or even the speculated quantum spin liquid state separating the PS and AF phases [50–52] could emerge from along such hypothesized tuning axes. Similar behavior has been observed in the bi-critical point of heavy fermion compound YbAgGe [57] with the specfic heat, the

Grüneisen parameter and the magnetocaloric effect, it will be of interest that the latter measurements can be performed here.

It is also interesting to note that phenomenon observed in this study shares the similarity with that seen in the pressurized Fe-based superconductor $Ca_{0.73}La_{0.27}FeAs_2$ [8], where a first-order transition of AF-SC phases is observed, and in that case, the two finite temperature phase boundaries (PM-AF and PM-SC phases) also meet at a bi-critical point. It is certain of significance to carry out the comparison on the two different materials, $SrCu_2(BO_3)_2$ and $Ca_{0.73}La_{0.27}FeAs_2$, by distinguishing the commonness and peculiarity, and find clues to understand the superconductivity in the proximity of AF phase. It is expected that our results will provide valuable experimental foundation and theoretical inspirations for eventually extending the paradigm of quantum phase transitions beyond the Landau-Ginzberg-Wilson.

## METHODS

The high-pressure heat capacity of the samples is derived from alternate-current (ac) calorimetry. In this technique, a small temperature oscillation ($\Delta T$) generated by a heater glued to one side of a sample is converted to an ac voltage signal that can be detected by a chromel-AuFe (0.07%) thermocouple fixed on the opposite side. The sample is loaded into a Teflon capsule filled with a mixture of glycerin-water (3:2), which can maintain the sample in a hydrostatic pressure environment. The ac calorimetry method adapted to high pressures is described in Refs [61, 62]. Pressure is determined by the pressure dependence of $T_c$ of Pb [63] that is placed together with the sample in the capsule. The detailed high-pressure heat capacity measurement platform is discussed in the SI.

## DATA AVAILABILITY

The data of this work is available upon reasonable requests.

* The three authors contributed equally to this work.
† zymeng@hku.hk
‡ llsun@iphy.ac.cn

## ACKNOWLEDGEMENTS

This work was supported by the National Key Research and Development Program of China (Grant No. 2021YFA1401800 and 2022YFA1403900, No. 2022YFA1403400, No. 2021YFA1400400), the NSF of China (Grant Numbers Grants No. U2032214, 12122414, 12104487 and 12004419), and the Strategic Priority Research Program (B) of the Chinese Academy of Sciences (Grant No. XDB25000000, No. XDB33000000), the K. C. Wong Education Foundation (GJTD-2020-01). J. G. and S.C. are grateful for supports from the Youth Innovation Promotion Association of the CAS (2019008) and the China Postdoctoral Science Foundation (E0BK111). C. H., B.-B.C and Z. Y. M. acknowledge the support from the Research Grants Council (RGC) of Hong Kong Special Administrative Region (SAR) of China (Projects Nos. 17301420, 17301721, AoE/P-701/20, 17309822 and HKU C7037-22G), the ANR/RGC Joint Research Scheme sponsored by the RGC of Hong Kong SAR of China and French National Research Agency (Project No. A HKU703/22) and the HKU Seed Funding for Strategic Interdisciplinary Research "Many-body paradigm in quantum moiré material research".

Corresponding authors
Correspondence to Zi Yang Meng (zymeng@hku.hk) or Liling Sun (llsun@iphy.ac.cn).